\documentclass{procDDs}
\usepackage{color}

\title{On semi-classical spectral series for an atom in a periodic polarized electric field}

\author{A.IFA}
{Universit\'e de Tunis El-Manar, D\'{e}partement de  Math\'{e}matiques, 1091 Tunis, Tunisia, \& Shaqra University, Saudi Arabia}                 
{wahebifa@su.edu.sa}

\author{H.LOUATI}
{Universit\'e de Tunis El-Manar, D\'{e}partement de  Math\'{e}matiques, 1091 Tunis, Tunisia, \& Northern Borders University, Saudi Arabia} 
{louatihanen42@yahoo.fr}
\coauthor{\bf{M.ROULEUX}}                       
{Aix Marseille Univ, Universit\'e de Toulon, CNRS, CPT, Marseille, France}               
{rouleux@univ-tln.fr}                                               
\def\Const{\mathop{\rm Const.}\nolimits}

\def\e{\mathop{\rm \varepsilon}\nolimits}
\def\Hes{\mathop{\rm Hess}\nolimits}

\def\im{\mathop{\rm Im}\nolimits}

\def\re{\mathop{\rm Re}\nolimits}
\def\rot{\mathop{\rm rot}\nolimits}

\def\Tr{\mathop{\rm Tr}\nolimits}

\def\WKB{\mathop{\rm WKB}\nolimits}
\begin {document}

\newtheorem{prop}{Proposition}

\newtheorem{rem}{Remark}
\maketitle
\index{Author1, I.I.}                              
\index{Author2, I.I.}                              
\index{Coauthor, I.I.}                             %

\begin{abstract}
In this report we present preliminary results about the tunneling problem for a magnetic Schr\"odinger operator.
As a motivation we consider the 3-D time-dependent Schr\"odinger operator 
$H(t)=-h^2\Delta+V+E(t)\cdot x$ where $V$ is a radial potential and $E(t)$ 
a circularly polarized field with uniform frequency $\omega$. 
The quantum monodromy operator (QMO)
that takes the system through a complete period $T=2\pi/\omega$, turns out to be unitarily equivalent to
$e^{iTP_A(x,hD_x)/h}$, where
$P_A(x,hD_x))$ identifies with a magnetic Schr\"odinger operator.
When $V$ is sufficiently confining, $P_A(x,hD_x))$ presents a double magnetic well.
Then we construct its semi-classical ground state
and examine the splitting between its two first eigenvalues.
\end{abstract}

\section{Introduction}

The general motivation is to 
describe some (semi-classical) spectrum for the quantum monodromy operator (QMO)  associated with a periodic time dependent  
self-adjoint $N$-body Schr\"odinger operator $H(t)$
on $L^2({\bf R}^{3N})$. Recall QMO is the unitary operator that evolves the physical system through one period.
This spectrum is the same as for Floquet Hamiltonian $\widehat H=D_t+H(t)$ on $L^2({\bf R}^{3N}_x\times{\bf R}_t)$, 
see e.g. \cite{Sch-Mo}.
Scattering theory for Floquet Hamiltonian leads to important properties, such as local decay (in time) for the probability density.
Bound states are also of interest, e.g. regarding tunneling properties.
A.Tip \cite{Tip} has investigated the case where
$H(t)$ is the Hamiltonian for an atom with $N$ particles of charge $e_j$ and mass $m_j$ 
($e_0=N$ being the charge of the nucleus of mass $m_0$) 
subject to the circularly polarized periodic electric field 
$E_1(t)=(\sin\omega t,-\cos\omega t,0)\in{\bf R}^3$. 
Let also $V$ (rotationally invariant in ${\bf R}^{3N}$) be the sum of all Coulomb potentials between the particles of charge $e_j$.
Neglecting spin effects, we have
\begin{equation*}
  H_1(t)=\sum_{j=0}^{N-1}(2m_j)^{-1}p_j^2+V(x)-\sum_{j=0}^{N-1}e_j x_j\cdot E_1(t)
\end{equation*}
Time-evolution $U_1(t,s)$ corresponding to $H_1(t)$ is well defined, and QMO is $U_1(T+s,s)$ (independent of $s\in{\bf R}$).
An equivalent formulation can be given in terms of 
\begin{equation*}
H_2(t)=\sum_{j=0}^{N-1}(2m_j)^{-1}\bigl(p_j-e_jA_2(t)\bigr)+V(x)
  \end{equation*}
with $E_1(t)=-\partial_tA_2(t)$, and 
the corresponding time-evolutions $U_j(t,s)$, $j=1,2$ are related by a unitary operator $X(t)$. 
One of the main observations in \cite{Tip} is that the group of unitary operators $R(t)=e^{i\omega tL_3}$, $L_3$
being the vertical component of the angular momentum (tensor product over the $N$ particles), takes $A_2(t)$ into a constant
magnetic potential $A_0=({1\over\omega},0,0)$, i.e.
\begin{equation*}
  R(t)H_2(t)R(t)^{-1}=\sum_j(2m_j)^{-1}(p_j-e_j A_0)^2+V(x)
  \end{equation*}
and 
\begin{equation*}
  U_2(t,s)=e^{-i\omega t L_3}e^{i(t-s)P_A}e^{i\omega s L_3}
  \end{equation*}
where $P_A$ is the time-independent self-adjoint operator
\begin{equation*}
P_A=\sum_{j=0}^{N-1}(2m_j)^{-1}(p_j-e_j A_0)^2+V(x)-\omega L_3
  \end{equation*}
Quantum monodromy operator $U_2(s+T,s)$ over the period $T=2\pi/\omega$, is of the form
\begin{equation*}
  U_2(s+T,s)=e^{-i\omega s L_3}e^{iTP_A}e^{i\omega s L_3}
  \end{equation*}
This relates the (discrete) spectra of $U_2(s+T,s)$ and $P_A$ (either real spectrum or resonances),
in the sense that if $E$ is an eigenvalue of $P_A$, then $F=e^{iTE}$ is an eigenvalue of $U_2(s+T,s)$. 
For simplicity we assume henceforth there is only one heavy nucleus and one electron $x\in{\bf R}^3$, 
and introduce a semi-classical parameter $h$
so that $L_3=x_1hD_{x_2}-x_2hD_{x_1}$, and $P_A$ on $L^2({\bf R}^3)$
takes the form 
$P_A(x,hD_x)=(hD_x-\nu A_0)^2+V(x)-\omega L_3$,
where $\nu$ is a coupling constant. 
In case $V$ is Coulomb potential, the complex dilation theory of $P_A$ set up in \cite{Tip} gives an insight into
the resonant spectrum of $P_A$. 
Here we still assume $V$ to be rotation invariant, but
allow for non Coulomb (confining) potentials, so that $P_A$ has discrete, real spectrum.

\section{Hamiltonians}

The classical Hamiltonian is
\begin{equation}\label{1.1}
  p_A(x,\xi)=(\xi_1-{\nu\over\omega})^2+\xi_2^2+\xi_3^2+V(|x|)-\omega(x_1\xi_2-x_2\xi_1)
  \end{equation}
and we may rewrite $P_A$ as a magnetic Schr\"odinger operator. Namely, after the affine change of coordinates
$x={2\nu\over\omega^2}y+(0,{2\nu\over\omega^2},0)$, and the substitution
$W(y)={4\nu^2\over\omega^4}\bigl(V(|x(y)|)-{\lambda^2\over\omega^2}(y_1^2+y_2^2)\bigr)$, 
we get $p_A(x,\xi)={\omega^4\over4\nu^2}p'_A(y,\eta)+{\nu^2\over\omega^2}$, where
\begin{equation}\label{1.2}
\begin{aligned}
  p'_A&(y,\eta)=(\eta_1+{2\nu^2\over\omega^3}y_2)^2+(\eta_2-{2\nu^2\over\omega^3}y_1)^2+\eta_3^2\cr
  &+W(y)=(\eta-\omega'A(y))^2+W(y)
  \end{aligned}
  \end{equation}
Note that the corresponding operator $P'_A(y,hD_y)$ commutes with symmetry $y_3\mapsto -y_3$. 
General spectral properties of magnetic Schr\"odinger operators are well-known, see e.g. \cite{He}, in particular when $W(y)\to+\infty$
as $y\to\infty$,
$P'_A(y,hD_y)$ has only discrete spectrum.  

Discrete semi-classical spectrum of $P_A$ near some energy level $E$ is generally associated with local minima of the Hamiltonian. 
Among these minima it is easy to find those
which are the critical points of $r\mapsto V(r)$, namely
\begin{equation*}
  \rho_0^\pm=(x_0,\xi_0)^\pm=\bigl((0,2\nu\omega^{-2},\pm\sqrt{r_0^2-4\nu^2\omega^{-4}}), 0\bigr)
  \end{equation*}
with $V(r_0)=E-\omega^2\nu^{-2}$.
We will write as well $\rho_0^\pm=(y_0^\pm,\eta_0^\pm)=(y_0^\pm,0)$ in the new coordinates $y$.

So the $y$-projection of the critical points $y_0^\pm$ of $p'_A(y,\eta)$ are located on the $y_3$-axis, symmetric with respect to
the $(y_1,y_2)$ plane. At such a critical point,  $\eta_0^\pm=0$. 

For a suitable choice 
of parameters $(r_0,V''(r_0),\omega,\nu)$, it may happen that $\rho_0^\pm$ is an elliptic point even if $r_0$ is not a minimum of $V$.
However, to meet hypotheses of \cite{MaSo}, we assume that $E$ is the bottom of the spectrum of $P_A$, and thus $r_0$ is a global minimum of $V$.
To avoid multiple tunneling, we assume also that there is no other critical points of $p_A$ than $\rho_0^\pm$ at energy $E$.

Such elliptic fixed points $\rho^\pm_0$ of Hamiltonian $p_A$, will be called the {\it magnetic microlocal wells}.
We consider Dirichlet realization 
$P_A^{M^\pm}(x,hD_x)$
localized in some (large enough) neighborhoods $M^\pm\subset{\bf R}^3$ of $x_0^\pm$ (by symmetry $P_A^{M^\pm}(x,hD_x)$ 
are unitarily equivalent) and study their semi-classical spectrum in a $h$-neighborhood of $E$, see \cite{HeSj1}. 
Note that the spectral parameters for $P_A^{M^\pm}$ and ${P'_A}^{M^\pm}$ are related as follows:
\begin{equation}\label{1.3}
  \begin{aligned}
    &P_A^{M^\pm}(x,hD_x)u_\pm(x)=(E+\lambda_A(h))u_\pm(x) \cr
    &\Longleftrightarrow \ {P'_A}^{M^\pm}(y,hD_y)u'_\pm(y)=(E'+\lambda'_A(h))u'_\pm(y)
\end{aligned}
\end{equation}    
where $E'={4\nu^2\over\omega^4}V(r_0)$, $\lambda'_A(h)={4\nu^2\over\omega^4}\lambda_A(h)$,
and $u'_\pm(y)=u_\pm(x)$ we will denote for short again by $u_\pm(y)$.
We know that the first eigenvalue $\lambda_A(h)$ of $P_A^{M^\pm}(x,hD_x)-E$ is non degenerate. 
By tunneling, this will gives raise to splitted eigenvalues
$E_0(h)<E_1(h)$ for $P_A(x,hD_x)$ exponentially close to $\lambda_A(h)$.
Thus our main goal consists in solving the rather standard problems: (1) Find semi-classical asymptotics of $\lambda_A(h)$, $h\to0$. 
(2) Estimate the splitting $E_1(h)-E_0(h)$ (tunneling between $\rho_0^\pm$). Point (1) results directly from the general
microlocal arguments of \cite{MaSo}
near $\rho_0^\pm$ applied to $P_A$, that we recall here.
Point (2) relies more specifically on the fact that $P_A$ (or $P'_A$) is a magnetic Schr\"odinger operator.
Both use complexification of time $t\mapsto it$ for Hamilton equations in the ``classically forbidden region''.

\section{Microlocal approach: WKB expansions near the microlocal magnetic well}

We work with $P_A$ rather than the rescaled operator $P'_A$, and since our constructions will hold  in a small neighborhood of $x_0=x_0^+$,
$x_0\in M=M^+$ say, 
we shall write sometimes $P_A$ instead of $P_A^{M^+}(x,hD_x)$.
At an elliptic fixed point
Floquet exponents, i.e. eigenvalues of the fundamental matrix $F_{p_A}={\cal J} \Hes p_A(\rho_0^\pm)$,  are purely imaginary. 
Their square $\mu'^2$
are the roots of a 3:rd degree polynomial with real coefficients. It turns out that we can find an open set of parameters $(r_0,V''(r_0),\omega,\nu)$ 
such that $r_0\omega^2\nu^{-1}\geq2$ and ${\mu'}_1^2<{\mu'}_2^2<{\mu'}_3^2<0$.  
When these conditions are met,  
applying a linear canonical transformation $\kappa_A$, and setting the reference energy to $E'=0$ to simplify notations, which amounts to set $V(r_0)=0$,
$p_A$ takes the form, with $\mu'_j=i\mu_j$, $\mu_j>0$
\begin{equation}\label{2.1}
  p_A(x,\xi)=\sum_{j=1}^3\mu_j(x_j^2+\xi_j^2)/2+{\cal O}(|x,\xi|^3)
  \end{equation}
in local coordinates $(x,\xi)$ vanishing at $\rho_0^\pm$ (but not the same as the original coordinates of the problem). It is important to
note that, although $\kappa_A$ changes drastically the geometry of the phase-space, we still get informations on
$u'_\pm(y)$, see \cite{MaSo}, Remark 5.2 and below. Namely, $\kappa_A$ does not generate additional caustics in the region of interest.

Examine first WKB constructions and decay properties of the normalized quasi-mode associated with $\lambda_A(h)$
by microlocal technics and the method of complex deformations \cite{HeSj1}, \cite{HeSj2}, \cite{He}, \cite{Ma2}. We follow here \cite{MaSo}.  

Consider the global FBI transform (with $c(h)$ a normalization constant)
\begin{equation}\label{2.2}
  Tu(x,\xi;h)=c(h)\int_{{\bf R}^3} e^{i(x-y)\xi/h-(x-y)^2/2h} u(y)\,dy
  \end{equation}
that maps isometrically $L^2({\bf R}^3)$ into $L^2({\bf R}^6)$. 
It is associated with the transformation $\kappa_T:T^*{\bf R}^3\to\Lambda=\{\zeta=i\im z\}$, $(x,\xi)\mapsto 
(z,\zeta)=(x-i\xi,-i\xi)$, such that $-\im(dz\wedge d\zeta)=dx\wedge\,d\xi$, $\re(dz\wedge d\zeta)=0$ on $\Lambda$. 
The pluri-subharmonic (pl.s.h) weight defining $\Lambda$, is $-\phi_0(z,\overline z)=(\im z)^2$, namely
$\Lambda=\{\zeta={\partial\phi_0\over\partial z}\}$. 
Transformation $T$ intertwines $P_A$ with $\widetilde P_A(x,\xi,hD_x,hD_\xi)$ whose Weyl symbol is given by 
$\widetilde p_A(x,\xi;x^*,\xi^*)=p_A^w(x-\xi^*,x^*)$,
so that the eigenvalue equation becomes, with $u_T=Tu$ and $u=u_+$ as in (\ref{1.3})
\begin{equation}\label{2.3}
  (\widetilde P_A(x,\xi,hD_x,hD_\xi)-\lambda_A(h))u_T(x,\xi;h)\sim0
  \end{equation}
which we will solve by WKB method, by looking for $u_T$ of the form $u_T(z;h)\sim e^{-\xi^2/2h-\varphi(z)/h}a(z;h)$.
The reduced operator $\Psi(x,\xi,hD_x,hD_\xi)=e^{\xi^2/2h+\varphi(z)/h}\widetilde P_A(x,\xi,hD_x,hD_\xi) e^{-\xi^2/2h-\varphi(z)/h}$
acting (formally) on symbols $a(z;h)$ is indeed a $h$-PDO $\Psi(z,hD_z;h)$ with symbol
$\sigma_\Psi(z,\zeta;h)\sim \sigma_0(z,\zeta)+h\sigma_1(z,\zeta)+\cdots$,
and $\sigma_0(z,\zeta)=-p_A(z-\zeta,i\zeta)$.

\begin{rem}
We could also use the standard unitary Bargman transform \cite{Sj}
$T_0:L^2({\bf R}^3)\to H_\Phi({\bf C}^3)$, space of holomorphic functions square integrable with respect to the weight $\Phi(z)=|x|^2/4$.
It is somewhat more natural in this context, but we found it convenient to stick to notations of \cite{MaSo}, \cite{Ma3}.
\end{rem}

Performing another ``absolute'' metaplectic complex transformation $T_2$ (independent of all parameters), associated
with a complex canonical transformation
$\kappa_2$, 
we can arrange so that $\sigma_0$ has the quadratic approximation 
$\sigma_{02}(z,\zeta)=\sum_{j=1}^32\mu_jz_j\zeta_j$ near the magnetic microlocal well. We still denote by $u_T$ the corresponding quasi-mode.
In fact $T_2$ could be avoided if we replace $T$ by $T_0$. 

We solve Hamilton-Jacobi (HJ) equation $\sigma_{02}(z,\partial_z\varphi_0(z)=0$ by $\varphi_0(z)={1\over4}z^2$.
This gives the complex ``outgoing'' Lagrangian manifold
$\Lambda^0_+=\{\zeta=\partial_z\varphi_0(z)\}$, and also the ``incoming'' one $\Lambda^0_-=\{\zeta=-\partial_z\varphi_0(z)\}$.
By the stable-unstable manifold Theorem this carries to $\sigma_0(z,\zeta)$, thus we can solve HJ equation with holomorphic phase function
$\varphi(z)=\varphi_0(z)+{\cal O}(|z|^3)$ defining $\Lambda_\pm$. Continuing this way, we can also
solve the transport equations with an analytic symbol $a(z;h)$. Altogether we get a WKB solution of $(\Psi(z,hD_z;h)-\lambda_A(h))a(z;h)\sim 0$
microlocally near 
$\rho_0=\rho_0^+$, such that
\begin{equation}\label{2.4}
  u_T(z,h)\sim e^{-\xi^2/2h-\varphi(z)/h}a(z;h)
  \end{equation}
Here $\sim$ means modulo a remainder term ${\cal O}(e^{-1/Ch})$. Applying classical variational principles, this is justified  
by Agmon type estimates, which say how close the microlocal solution approaches the actual
eigenfunction. To this end, we use the method of non characteristic deformations. Namely in some neighborhood $\Omega$ of (0,0)
where microlocal solution (\ref{2.4}) holds
(i.e. before the occurence of focal points on $\Lambda_\pm$)
there is a deformation of $-\phi_0$ to a family of pl.s.h. weights 
$-\phi_t(z,\overline z)$,  such that the R-Lagrangian manifold 
$\Lambda_t=\exp  tH_{q}(\Lambda)$ is transverse to
$\{z=0\}$ at (0,0), and hence of the form
$\zeta={\partial\phi_t\over\partial z}$. We check that $\phi_t(z,\overline z)\to2\re \varphi(z)$ as $t\to\infty$.
Moreover $\sigma_0$ remains elliptic outside (0,0)
along these deformations, in the sense $-\sigma_0(z,\zeta)|_{\Lambda_t}\geq \Const
(|z|^2+|\zeta|^2)$ for all $t$, and $\phi_t$ satisfies the eikonal equation
$\partial_t\phi_t(z,\overline z)=-2\sigma_0(z,\partial_t\phi_t)$. We then modify $\phi_t$ to another weight $\psi_t$ such that $\psi_t<2\re \varphi$
near $\partial\Omega$, where WKB constructions of type (\ref{2.4}) fail to exist, and get the following Agmon estimate:
If $u$ is the 1:st normalized eigenfunction of $P_A^{M}$ then for any compact set $K\subset\Omega$, there exists $\e >0$ and $h_{\e }>0$,
such that uniformly for $z=x-i\xi\in K$, $0<h<h_{\e }$ we have
\begin{equation}\label{2.5}
  \begin{aligned}
    e&^{\xi^2/2h+\varphi(x-i\xi)/h}(Tu_+(x,\xi;h)-u_T(x,\xi;h))=\cr
    &{\cal O}(e^{-\e /h})
    \end{aligned}
  \end{equation} 
Altogether, this gives the asymptotics for the first eigenvalue $\lambda_A(h)$ for $P_A^{M^+}(x,hD_x)$
\begin{equation}\label{2.6}
  \lambda_A(h)=h\Tr ^+(F_{p_A})+{\cal O}(h^2)
  \end{equation}
where $\Tr ^+(F_{p_A})=\sum_{j=1}^3 \mu_j$. This gives also the asymptotics of $Tu_+$ in $\Omega\subset M$. 

The main drawback is the poor control of the neighborhood of $x_0$ where this asymptotics holds
for $u=T^{-1}u_T$ (left inverse), because 
$\kappa_A$ is not merely implied by a change of variables in the $x$-space.
Nevertheless since $\kappa_T^{-1}(\widetilde\Lambda_+)$, $\widetilde\Lambda_+=\{\zeta=i\partial_z\varphi(z)\}$
is a strictly positive Lagrangian manifold at (0,0) (also after applying $\kappa_2$),
the same holds for $\kappa_A^{-1}\circ\kappa_T^{-1}(\widetilde\Lambda_+)$ (in the initial canonical variables),
and so $\Lambda_+$ has no focal points near $\rho_0$. So Agmon estimate (\ref{2.5}) again justifies
the asymptotics of $u$ of $P_A^{M}(x,hD_x)$
in some (real) neighborhood $\Omega_R$ of $x_0$. 

Of course by symmetry $y_3\mapsto -y_3$ everything carries to $M^-$. To proceed we need to extend WKB constructions,
when $\omega'$ is small enough, 
near minimal geodesics for Agmon metric associated with the Hamiltonian without a magnetic field, .

\section{Geometry and quantization of magnetic Hamiltonians}

We work here with the rescaled operator $P'_A(y,hD_y)=(P'_A)^M(y,hD_y)$ localized in $M=M^+$.
In Classical Mechanics the generalized momentum $\eta$ of a particle is related to its velocity by $v=\eta-A(y)$.
The magnetic symplectic 2-form is
\begin{equation*}
  \sigma_A=dv\wedge dy=d\eta\wedge dy+dA(y)=\sigma+dA(y)
  \end{equation*}
Consider first the kinetic term $K_0(y,\eta)=|v|^2=(\eta-A(y))^2$. We know \cite{Iv1}, \cite{RaVu} that if $dA(y)\neq0$, then 
$\Sigma=\{K_0=0\}=\{dK_0=0\}$ is a 3-D submanifold of $T^*{\bf R}^3$, and $\Sigma\cap\Sigma^\sigma$ ($\Sigma^\sigma$ denotes orthogonal complement
with respect to 
the symplectic 2-form $\sigma$) is the Hamilton vector field for $\sum_{j=1}^3F_j(y)(\eta_j-A_j(y)$ where $F_j(y)$ are the components 
of the vector intensity of the magnetic field $B(y)$. In the present case $\Sigma\cap\Sigma^\sigma={\bf R}{\partial\over\partial y_3}$. 
(Note that in the 2-D case, $\Sigma$ is just a 2-D symplectic manifold).
Many examples of integrable magnetic Hamiltonians are provided in \cite{Iv2}. 

As in the case of Schr\"odinger operator without a magnetic field, we may define complex branches of the energy surface by considering 
imaginary times. They glue along $\Sigma$. For Hamiltonian (\ref{1.1}) with a potential $V$ sufficiently confining,
the ``magnetic classically allowed region'' (MCAR) is a domain in ${\bf R}^3$
with 2 connected components
in $\pm y_3>0$ bounded by the closed 2-D manifolds defined by
$\pi_y\bigl(\Sigma\cap\{p'_A(y,\eta)=E'\}\bigr)$, where $\pi_y:T^*{\bf R}^3\to{\bf R}^3$ is the natural projection,
and the ``magnetic classically forbidden region'' (MCFR) its complement.
For small $\omega'$, the MCAR can be viewed as a deformation of the ``classically allowed region'' (CAR)
for the Hamiltonian $P'_0(y,hD_y)=(hD_y)^2+W(y)$
without a magnetic field.

We stress that in the MCFR we make the substitution
$v\mapsto iv$  instead of $\xi\mapsto i\xi$ as in \cite{HeSj2}. By this transformation
the Hamiltonian remains real and gauge invariant. This is consistent with the following example of operator on $L^2({\bf R}^2)$
\begin{equation}\label{DA}
  H(y,hD_y)=(hD_{y_1})^2+(hD_{y_2}-by_1)^2+V_1(y_1)+\omega^2y_2^2
  \end{equation}
which is unitarily equivalent (after partial Fourier transform with respect to $x_2$)
to the Schr\"odinger operator without a magnetic field $H'(x,hD_x)=(hD_{x_1})^2+(hD_{x_2})^2+(\omega x_2-bx_1)^2+V_1(x_1)$
(see \cite{BDN}, \cite{DA}).
It turns out that our computations lead (up to the present accuracy)
to the same quantities as \cite{HeSj2}, but without factor $i$ at some places, see e.g. Proposition 1
below. 

\begin{rem}
  Quantization of the kinetic term with a linear $A(y)=(-y_1,0,0)$ (Landau gauge) was considered first in \cite{LaLi}, p.496, leading to 
  Landau levels with infinite degeneracy. In this case, the $y$-projection of classical trajectories consist in a family of helices,
  with arbitrary center. The generalized wave-functions are expressed in term of a family of Hermite functions oscillating in the $y_1$-direction
  around the center. This gives a hint at understanding the case of magnetic Schr\"odinger operator (MSO)
$P'_A(y,hD_y)$~: degeneracy is lifted by the potential, that fixes somehow the center of the helix. 
\end{rem}

Notice that the potential well $U_{E'(h)}$ for $p'_0(y,\eta)=\eta^2+W(y)$ at energy $E'(h)={4\nu^2\over\omega^4}(V(r_0)+\lambda_A(h))$ with 2 connected
components $U\pm_{E'(h)}$ is given by
\begin{equation}\label{3.1}
  \begin{aligned}
    U_{E'(h)}&^\pm=\{y: V(|x(y)|)-{\nu^2\over\omega^2}(y_1^2+y_2^2)\leq \cr
    &V(r_0)+\lambda_A(h), \ \pm y_3>0\}
    \end{aligned}
  \end{equation}
Thus $y_0^\pm$ (the ``center'' of each well) verifies $y_0^\pm\in U_{E'}^\pm$,  and
$\partial U_{E'(h)}^\pm\subset M^\pm$
if $r\mapsto V(r)$ grows fast enough at infinity and near 0, so that in particular $W|_{y_3=0}>0$.
Under these conditions, by the hyperplane symmetry, $p'_0$ has two potential wells $U^\pm_{E'(h)}$ localized near $y_0=y_0^\pm$.
Conditions on Floquet exponents (see Sect.3) ensure that $W$ has nondegenerate minima at $y_0^\pm$. 

Taking Weyl quantization of $p'_A(x,\xi)=(\eta-\omega'A(y))^2+W(y)$, where $A(y)=(-y_2,y_1,0)$ (symmetric gauge) and $\omega'={2\nu^2\over\omega^3}$,
we get the magnetic Schr\"odinger Hamiltonian
\begin{equation}\label{3.3}
\begin{aligned}
P'_A&(y,hD_y)u(y;h)=(2\pi h)^{-3}\int_{{\bf R}^6} \cr
&e^{i(y-z)\eta/h}p'_A({y+z\over2},\eta)u(z)\,dz\,d\eta
\end{aligned}
\end{equation}
and by a change of variables
\begin{equation}\label{3.4}
\begin{aligned}
  P'_A&(y,hD_y)u(y;h)=(2\pi h)^{-3}\int_{{\bf R}^6} \cr
  &e^{i(y-z)\bigl(\eta+A\bigl({y+z\over2}\bigr)\bigr)/h}p'_0({y+z\over2},\xi)u(z)\,dz\,d\eta
  \end{aligned}
  \end{equation}

\begin{rem}
For non linear magnetic potentials $A(y)$ 
we need to modify this quantization, since it is no longer gauge invariant, see \cite{ManPu}.  
There are far-reaching generalisation of such $h$-PDO's, in particular Berezin-Toeplitz operators, see \cite{Ch1}, \cite{Ch2}, \cite{Hi}. 
\end{rem}

\section{WKB constructions for $P'_0$ and $P'_A$}

We will somehow relate the quasi-modes of $P'_0$ and $P'_A$ in $M^\pm$, outside $U^\pm_{E'(h)}$. Since we are looking for rather rough estimates
on the eigenvalue splitting, 
it is sufficient here to search for WKB solutions of 
$\bigl(P'_0(y,hD_y)-\lambda'_A(h)\bigr)w(y;h)\sim0$ in the classically forbidden region (CFR) $W(y)\geq E'(h)$,
although $P'_0$ and $P'_A$
do not have the same ground state, see e.g. the discussion in \cite{He}, Sect.7.2.

First we recall some known facts on tunneling for Schr\"odinger operators without a magnetic field.
Let $S_{E'(h)}^+(y)$ be Agmon distance from $U^+_{E'(h)}$ to $y$
for the Riemannian conformal metric $\bigl(W(y)-E'(h))\bigr)_+\,dy^2$, $E'(h)={4\nu^2\over\omega^4}(V(r_0)+\lambda_A(h))$
which vanishes on $U_{E'}$.
We know that the exponential decay of $w$ from the well
$U^+_{E'}$, say, is given roughly by $e^{-S_{E'}^+(y)/h}$,  see \cite{HeSj1}, \cite{Ma2}.  So we have to compute $w$
near minimal geodesics $\gamma_{E'(h)}$ between $\partial U^\pm_{E'(h)}$. Such (finitely many) minimal 
geodesics are also called {\it librations} \cite{BDN}, \cite{DA},
\cite{AnRo}. Within the required accuracy on tunneling rates, we could again replace $E'(h)$ by $E'$,
which amounts to replace the librations by the instanton between $U^\pm_{E'}=\{y_0^\pm\}$. 

Changing $t$ to $it$ we may assume that in $W(y)\geq E'(h)$ (CFR) $P'_0-E'(h)$ takes the form
$Q'_0(y,hD_y)+E'(h)=(hD_y)^2-W(y)+E'(h)$. 
Then the WKB solutions of $(Q'_0(y,hD_y)+E')w(y,h)=\lambda_A(h)w(y,h)$ are given by
\begin{equation}\label{4.1}
  w(y;h)=e^{-\phi(y,E')/h}b(y,E';h)
  \end{equation}

In particular the phase function $\phi$ solves (locally) Hamilton-Jacobi equation 
$(\nabla\phi(y,E'))^2-W(y)+E'=0$ 
and the amplitude $b(y,E';h)=b_0(y,E';h)+hb_1(y,E';h)+\cdots$ the
transport equations. These WKB expansions can be justified by Agmon estimates near the
minimal geodesics $\gamma_{E'}$ as in \cite{Ma2}. 

Let now $Q'_A(y,hD_y)+E'(h)=(hD_y-\omega'A(y))^2-W(y)+E'(h)$ be the corresponding Hamiltonian in MCFR.
Recall that MCFR (with the magnetic field of strength $\omega'$) is a  deformation ${\cal O}(\omega')$
of the CFR (without the magnetic field). 

To solve $(Q'_A(y,hD_y)+E'(h))u(y;h)\sim0$, we try 
\begin{equation}\label{4.3}
  \begin{aligned}
  &u(y;h)=e^{-\psi(y,E')/h}b(y,E';h)w(y,E';h)=\cr
  &e^{-(\phi(y,E')+\psi(y;E'))/h}b(y,E';h)c(y,E';h)
\end{aligned}
\end{equation}
and replace $E'(h)$ by $E'$ as stated above. We find (omitting parameter $E'(h)$), using (\ref{3.3}) and (\ref{3.4}) 
\begin{equation}\label{4.4}
\begin{aligned}
  &(Q'_A(y,hD_y)+E')u(y;h)=(2\pi h)^{-3}\int\int\cr
  &\exp \bigl[i\bigl(
(y-z)(\eta+\omega'A({y+z\over2}))+\phi(z)+\psi(z)\bigr)/h\bigr]\cr
&\bigl(q'_0({y+z\over2},\eta)+E'\bigr)b(z,E';h)c(z,E';h)\,dz\,d\eta
\end{aligned}
\end{equation}
(we have identified the operator with its Weyl symbol). Applying asymptotic stationary phase yields eikonal equation
\begin{equation}\label{4.5}
  (-\omega'A(y)+\phi'(y)+\psi'(y))^2-W(y)+E'=0
  \end{equation}
which is Hamilton-Jacobi equation for Hamiltonian
$(\eta-\omega'A(y)+\phi'(y))^2+W(y)+E'$. By a stability argument (\ref{4.5}) has again a unique solution near $\gamma_{E'}$, and
can be solved perturbatively in $\omega'>0$ small enough.

Namely, look for $\psi(y)=\omega'\psi_0(y)+\omega'^2\widetilde\psi_1(y,\omega')$. 
Substituting into (\ref{4.5}) we get at first order in $\omega'$:
$2\nabla\phi(y)\nabla\psi_0(y)=2\langle\nabla\phi(y),A(y)\rangle$
which is a transport equation along the integral curves of Hamilton vector field $H_{Q'_0}$. Using Hamilton Eq. for $Q'_0$ it can be written  as
${d\over dt}\psi_0(y(t))=2\bigl(-y_2{\partial\phi\over\partial y_1}+y_1{\partial\phi\over\partial y_2}\bigr)$
or in cylindrical coordinates $y_1=r\cos\theta,y_2=r\sin\theta,y_3=y_3$, 
${d\over dt}\psi_0(x(t))=2{\partial\phi\over\partial\theta}(x(t))$
which can be integrated as (see also \cite{HeSj2}, Eq.$(2.27)_1$)
$\psi_0(t)=\psi_0|_{{\cal H}_*}+2\int_0^t{\partial\phi\over\partial\theta}(x(s))\,ds$. 
Higher order approximations in $\omega'$ are obtained similarly. In particular 
setting $\widetilde\psi_1=\psi_1+\omega\widetilde\psi_2$, we find 
\begin{equation*}
  \begin{aligned}
    2&\nabla\phi(y)\nabla\psi'_1(y)=-(\omega' y)^2-{\psi'_0}^2+\cr
    &2\langle\ A(y),\psi'_0\rangle=-|\psi'_0-A(y)|^2
    \end{aligned}
  \end{equation*}
or (see also Eq. $(2.27)_2$ and Lemma 3.9 of \cite{HeSj2}),
${d\over dt}\psi_1(x(t))=-\int |\psi'_0(x(s))-A(x(s))|^2\,ds$ 
The same type of arguments holds for transport equations defining $c(y,E';h)$. 

Note that in dimension 1, the leading order term of the symbol $b(y,E';h)c(y,E';h)$ would assume the familiar form
$b(y,E';h)c(y,E';h)=\Const \exp\bigl[-{1\over2}\int^y(W(z)-E')^{-1/2}\,dz\bigr]$, and we expect this formula to hold in the 3-D case as well.

\begin{rem}
  The computation above is based on representation (\ref{3.4}) in the real domain, can actually be carried over MCFR
  where the phase becomes complex (possibly purely imaginary), if we choose suitable branches of solutions with positive imaginary part.
  This can e.g. be achieved after performing a FBI transform of type (\ref{2.5}) which allows for complex phases.
 \end{rem} 

This eventually gives $u=u_{\WKB }$ outside some $\omega'$-neighborhood of CAR
$\{y: W(y)\leq E'(h)\}$ of the form $\widetilde\Omega^+_{\omega'}\cup\widetilde\Omega^-_{\omega'}$.

In fact CFR and MCFR may differ by ${\cal O}(\omega')$. 
To ensure that our WKB solutions overlap with the microlocal solutions obtained in Sect.3, we need to assume that
representation (\ref{2.4}) of the actual eigenfunction holds in a sufficiently large neighborhood (of size ${\cal O}(\omega')$)
of the CAR $\{y: W(y)\leq E'\}$ (without the magnetic field). This requires that we have a good control, sufficiently far away the microlocal wells,
of the pl.s.h. weights
$-\phi_t$ constructed in Sect.3. It should be instructive to consider the model (\ref{DA}) from this viewpoint.

Let us summarize our constructions in the following:

\begin{prop}: Assume $\widetilde\Omega^+_{\omega'}\cup\widetilde\Omega^-_{\omega'}\subset \Omega^+_R\cup\Omega^-_R$,
  where $\Omega^\pm_R$ have been defined at the end of Sect.2.
  Then there are WKB solutions $u_{\WKB }$ as in (\ref{4.3}) of $(P_A(x,hD_x)-E-\lambda_A(h))u_{\WKB }\sim0$ extending the quasi-modes
  $u^\pm$ of $P_A^{M^\pm}$ constructed
in (\ref{2.4}) along the (finite set) of minimal geodesics $\gamma_{E'}$ between $\partial U^\pm_{E'(h)}$. 
\end{prop}
As in the case of Schr\"odinger operator without a magnetic field, we can control the decay of eigenfunctions outside $U^\pm_{E'}$
by Agmon estimates.  We start with a rough weighted energy estimate.
Let ${\cal E}_a=\{y: d_{E'}(U_{E'}^+,y)+d_{E'}(U_{E'}^-,y)\leq S_{E'}+a\}\subset M^+\cup M^-$ for suitable $a>0$, and  
$\widetilde\Omega\subset{\bf R}^3$ large enough with $C^2$ boundary, such that $U^+_{E'}\subset\widetilde\Omega$, $U^-_{E'}\cap\widetilde\Omega=\emptyset$,
$\Gamma=\partial\Omega\cap{\cal E}_a\subset M^+\cap M^-$, see \cite{He},p.43. Let also $\Phi$ be a Lipschitz function on $\partial\widetilde\Omega$.
Then for all $u\in C^2(\overline\Omega;{\bf C})$, $u|_{\partial\widetilde\Omega}=0$, we recall from \cite{He},Prop.7.2.25
\begin{equation}\label{4.6}
\begin{aligned}
  &\int_{\widetilde\Omega} |(hD_x-A(x))(e^{\Phi/h}u)|^2\,dx+\cr
  &\int_{\widetilde\Omega}(W(x)-E'-|\Phi'(x)|^2)e^{2\Phi/h}|u|^2\,dx\cr
&=\re \int_{\widetilde\Omega} e^{2\Phi/h}(P'_A-E')u(x) \overline u(x)\,dx
\end{aligned}
\end{equation}
The weight $\Phi$ should of course be related with $\phi+\psi$, but the term $(W-E'-|\Phi'(x)|^2)$ in the second integral
refers rather to the eikonal equation
verified by $\phi$ alone, so instead of the decay observed in the WKB solution, we end up with this we would get in case $A(x)=0$.
So sharper estimates would require either to
assume analyticity of the potential as in \cite{HeSj2}, or 
to change formula (\ref{4.6}), to a formula verified by the ratio $\phi_A=u_A/u_0$ of ground states with and without a magnetic field.
Note that in case $\lambda_A(h)=\lambda_0(h)$, which is achieved in particular (Bohm-Aharonov vector potential)
when $\rot A=0$
in $\widetilde\Omega$, necessarily non simply connected,  and $\int_\gamma A(x)\,dx\in 2\pi h{\bf Z}$ for all closed path
$\gamma\subset\widetilde\Omega$, then
$(hD_x-A(x))\phi_A=0$. This is due to an identity by R.Lavine and M.O'Carroll, see \cite{He},p.93 and references therein,
and leads in particular that $\phi_A$
is just a ($x$-dependent) phase factor. So in this case there would be no additionnal decay due to the magnetic field.

\section{The gap formula and estimate of the splitting}. 

As in the case of Schr\"odinger operator without a magnetic field, we relate the splitting to an integral of $u_+{\partial u_-\over\partial y_3}$
over $\Gamma\subset\{y_3=0\}$ that bissects the classically forbidden region near the minimal geodesics. The gap formula is given by the interaction 
matrix using Stokes formula with the magnetic potential
\begin{equation}\label{4.8}
\begin{aligned}
  &w_{+-}=h^2\int_\Gamma \bigl(\overline {u_+}{\partial u_-\over\partial y_3}-u_-{\partial \overline {u_+}\over\partial y_3}\bigr)\,dS(y)+\cr
&h\int_\Gamma \bigl(\overline {u_+}(y)\langle A(y),{\partial\over\partial y_3}\rangle u_-(y)\cr
&-u_-(y)\langle A(y),{\partial\over\partial y_3}\rangle u_+(y)\bigr)\,dS(y)
\end{aligned}
\end{equation}

The second term vanish since $A_3(y)=0$, hence (\ref{4.8}) is identical to the gap formula without a  magnetic field.  
This altogether with (\ref{4.6}) yields an estimate for the splitting of the form
\begin{equation}\label{4.7}
  e^{-\bigl(S_{E'}+{\cal O}(\omega')\bigr)/h}\leq E_1(h)-E_0(h)\leq e^{-\bigl(S_{E'}-{\cal O}(\omega')\bigr)/h}
  \end{equation}
but getting a sharper exponent requires to modify (\ref{4.6}) as suggested above.
Precise Agmon estimates also depend on the number of these geodesics,
and on the structure of the eigenfunction $u_A$ of $P_A(y,hD_y)$ near $\partial U^\pm_{E'(h)}$, obtained in Sect.3. 

\section*{Acknowledgements} We thank B.Helffer for interesting discussions on Agmon estimates in the presence of a magnetic field,
and a referee for some useful references.


\begin{thebibliography}{99}

\bibitem{AnRo} Anikin, A.\, Rouleux, M.\, Multidimensional tunneling between potential wells at non degenerate minima.
Proceedings \textit{Days of Diffraction 2014}, p.\;17-22. Saint-Petersburg.

\bibitem{BDN} Br\"uning, J.,\, Dobrokhotov, S., \, Nekrasov, R. \, 2013, Splitting of lower energy levels in a quantum double well in a
magnetic field and
tunneling of wave packets in Nanowires. Theor. Math Phys 175, p.620--636.

\bibitem{Ch1} Charles, L.\, 2000, Aspects semi-classiques de la quantification g\'eometrique. \textit{Th\`ese, Universit\'e Paris IX-Dauphine}.

\bibitem{Ch2} Charles, L.\, 2006, Symbolic calculus for Toeplitz operators with half-form. \textit{J. Symplectic Geometry}, Vol.\;\textbf{4}(2),
  pp.\;171-198.
  
\bibitem{CFKS} Cycon, H.\, Froese, R.,\, Kirsch, W,\, Simon, B.\, 1987, \textit{Schr\"odinger Operators}. Springer.

\bibitem{DA} Dobrokhotov, S.,\, Anikin, A.\, 2014, Tunneling, Librations and Normal Forms in a Quantum Double Well with a Magnetic Field.
in {\it Nonlinear Physical Problems, Spectral Analysis, Stability and Bifurcations}, Oleg N.Kirillov, Dmitry E. Pelinovsky, {\it Eds.}
Series Editor Noel Challamel, John Wiley \& Sons, pp.85--108.

\bibitem {He} Helffer, B.\, 1988, Semi-classical Analysis for the Schr\"odinger and applications. \textit{Lecture Notes in Math},
  Vol.\;\textbf{1336}, Springer.

\bibitem {He1} Helffer, B.\, 1994, On Spectral Theory for Schr\"odinger Operators with Magnetic Potentials. Adv. Studies in Pure Math. 23,
  1994, Spectral and Scattering Theory and Applications, pp.113-141
  
\bibitem{HeSj1} Helffer, B., Sj\"ostrand, J., 1984, Multiple wells in the semi-classical limit I, \textit{Comm. Part. Diff. Eqn.}, 
  Vol.\;\textbf{9}(4) pp.\;337-408.

\bibitem{HeSj2} Helffer, B., Sj\"ostrand, J., 1987, Effet tunnel pour l'\'equation de Schr\"odinger avec champ magn\'etique.
\textit{Ann. Scuola Norm. Sup. Pisa} Cl. Sci. \textbf{4}(14), No.4, pp.\;625--657.
1987. 

\bibitem{Hi} Hitrik, M., Asymptotic Bergman kernels in the analytic case, {\it this Conference}.

\bibitem{Iv1} Ivrii, V., 2019, \textit{Microlocal Analysis, Sharp Spectral Asymptotics and Applications}, Vol. I-V, Springer.  

\bibitem{Iv2} Ivrii, V., Magnetic Schrodinger operators: Geometry, Classical and Quantum Dynamics and Spectral Asymptotics.
S\'eminaire EDP (Polytechnique) (2006-2007), Expos\'e No. 16, 23 p. 

\bibitem{LaLi} Landau, L.\, Lifschitz, E.\, 1967, \textit{M\'ecanique Quantique}. Mir, Moscou.

\bibitem{ManPu} Mantoiu, M.\, Purice, R.\, 2004, The Magnetic Weyl Calculus. \textit{J. Math. Phys}, Vol.\;\textbf{1336}, pp.1394--1417.

\bibitem{Ma1} Martinez, A., 1987, Estimations de l'effet tunnel pour le double puits I. \textit{J. Math Pures et Appli.}, Vol.\; \textbf{66}, pp.195--215

\bibitem{Ma2} Martinez, A., 1988, Estimations de l'effet tunnel pour le double puits II. Etats hautement excit\'es.\textit{Bull. Soc. Math. France},  Vol.\; \textbf{116}(2)\, pp.199--219.

\bibitem{Ma3} Martinez, A., 2002, \textit{An Introduction to Semiclassical and Microlocal Analysis}. Springer.

\bibitem{MaSo} Martinez, A.\,  Sordoni, V., 1999. Microlocal WKB expansions. \textit{J.Funct.Anal.} Vol.\; \textbf{168}, pp.380--402.

\bibitem{RaVu} Raymond, N.\, Vu Ngoc, S., 2015, Geometry and Spectrum in 2D Magnetic Wells. \textit{Annales Institut
Fourier} Vol.\; \textbf{65}(1), pp.137--169.

\bibitem{Sch-Mo} Schach-M\"oller, J., 2000, 2-body short-range systems in a time-periodic electric field. \textit{Duke Math.J.}, Vol.\;\textbf{105}(1). 

\bibitem{Sj} Sj\"ostrand, J., 1982, Singularit\'es analytiques microlocales, \textit{Ast\'erisque} Vol.\; \textbf{95}, Soc. Math. France.

\bibitem{Tip} Tip, A, 1983, Atoms in circularly polarized fields: 
the dilation-analytic approach. \textit{J.Phys. A Math. Gen. Phys.} Vol.\; \textbf{16}, pp.3237--3259.

\end{thebibliography}
\end{document}